\documentclass[10pt, conference, compsocconf]{IEEEtran}
\ifCLASSINFOpdf
\else
\fi
\begin{document}
%
\title{Lost in translation: data integration tools meet the Semantic Web\\(experiences from the Ondex project)}


\author{\IEEEauthorblockN{Andrea Splendiani, Chris J Rawlings, Shao-Chih Kuo}
\IEEEauthorblockA{Biomathematics and  Bioinformatics Dept.\\
Rothamsted Research\\
Harpenden,\\ United Kingdom}
\and
\IEEEauthorblockN{Robert Stevens}
\IEEEauthorblockA{School of Computing Science\\
University of Manchester\\
Manchester,\\ United Kingdom}
\and
\IEEEauthorblockN{Phillip Lord}
\IEEEauthorblockA{School of Computing Science\\
Newcastle University\\
Newcastle upon Tyne,\\ United Kingdom}
}


%


\maketitle

\begin{abstract}
More information is now being published in machine processable form on
the web and, as \emph{de-facto} distributed knowledge bases are
materializing, partly encouraged by the vision of the Semantic Web, the focus is
shifting from the publication of this information to its consumption.
Platforms for data integration, visualization and analysis that are
based on a graph representation of information appear first candidates
to be consumers of web-based information that is readily expressible as
graphs.  The question is whether the adoption of these platforms to
information available on the Semantic Web requires some adaptation of
their data structures and semantics.  Ondex is a network-based data
integration, analysis and visualization platform which has been
developed in a Life Sciences context. A number of features, including
semantic annotation via ontologies and an attention to provenance and
evidence, make this an ideal candidate to consume Semantic Web
information, as well as a prototype for the application of network
analysis tools in this context.  By analyzing the Ondex data structure
and its usage, we have found a set of discrepancies and errors arising
from the semantic mismatch between a procedural approach to network
analysis and the implications of a web-based representation of
information.  We report in the paper on the simple methodology that we
have adopted to conduct such analysis, and on issues that we have
found which may be relevant for a range of similar platforms.
\end{abstract}


%
\IEEEpeerreviewmaketitle

\section{Introduction}

In this paper we describe a simple methodology used to examine a graph
based data resource so that it can be transformed to a representation
suitable for the Semantic Web. Such simple methodologies are needed if
Semantic Web technologies are to be used as widely as possible. The
web has been a revolutionary technology to exchange and integrate
information represented in natural language that has enabled the
development of new means of communication and interaction. Now the web
is evolving into a platform that also supports the integration and
exchange of machine processable information. This platform has the
potential to enable radical new approaches in the way we make sense of
information. It has been the object of active research, from the
Semantic Web \cite{od:CIT1} to its more recent development as Linked
Data \cite{od:CIT2}.

An increasing number of resources are now available on the Semantic
Web, either exporting their information in standard languages such as
the Resource Description Framework (RDF \cite{od:CIT4}), or directly
providing information servers that respond to standard query protocols
(SPARQL \cite{od:CIT5}). In addition, a number of key players are
committing to either publish their information on the Semantic Web, or
to support some related forms of structured knowledge publication and
consumption via the web, including national governments (UK
\cite{od:CIT6}, US\cite{od:CIT7}) and leading enterprises (Facebook
\cite{od:CIT8}, Google \cite{od:CIT9}, Yahoo \cite{od:CIT10}).

The availability of these information resources is complemented
by the increasing number of tools and web systems that natively
support the creation of information that is ported to a Semantic Web
framework. We cite as examples, tools such
CouchDB (couchdb.apache.org) and Neo4j (neo4jorg), web resources such
as Freebase (freebase.com). Refer to NoSQL
\cite{od:CIT12} for further details.

Now that `a' \emph{de-facto} Semantic Web is a reality, it is time to consider how it can be exploited and which software tools are needed to reap benefits from it. As the Semantic Web (as well as the traditional web), are founded on a graph-based representation of information, tools and methods for the analysis, manipulation and visualization of graphs are first candidates for this purpose, and we are witnessing the first examples in this direction, such as Gremlin (gremlin.tinkerprop.com) or RelFinder \cite{od:CIT16}. Most of these tools, however, have been developed following assumptions that do not necessarily apply to a web based representation of information and both their information engineering approach, and their usage need to be adapted for the Semantic Web context.
This is particularly true for several tools which have been developed in the domain of Life Sciences for the analysis of biological networks \cite{od:CIT16b}, some of which (e.g.:Ondex\cite{od:CIT17}, Cytoscape\cite{od:CIT19}) are inherently domain independent.
It should be noted that the Life Sciences present a number of information related issues that makes the Semantic Web an ideal solution, and computational biologists and bioinformaticians have been among the most enthusiastic adopters of these technologies \cite{od:CIT18}. 
The size of the user community and potential impact of the Semantic Web in the Life Sciences makes it a most attractive domain to deploy Semantic Web graph-based analysis and visualization tools (cfr. RDFScape \cite{od:CIT20}).

When considering software which could be made available for users of the Semantic Web, we wish to ask how we can evaluate if the usage of such tools is inconsistent with the principles of representation in the Semantic Web and what are the main aspects of the information engineering in these applications that may need to be
adapted to ameliorate any conflicts?

To answer these questions, we have focused on the Ondex data integration platform. 
Ondex is a data integration and analysis platform that has been developed, starting in 2005, for
research in systems biology and the Life Sciences in general, but which is inherently domain independent.  
Its information engineering design is based on a graph data structure and on the use of ontologies to characterize the graph entities.
Among other graph based tools developed in the Life Sciences, Ondex is unique for the precise semantic characterization of information and for its focus on graph-based data integration.
Furthermore, the information design of Ondex resembles that of RDF data in the Semantic Web, and a number of issues that users of Ondex deal with are essentially the same issues posed by information in a Semantic Web context (e.g., linking with provenance or evidence attached to the entities).  

We have developed a simple methodology to analyze the correspondence
between the intended semantics of these graph-based tools, and
the semantics resulting from their usage on web-data. This method was
designed with a view to adapting Ondex to work within the Semantic Web and to learn lessons
that would inform users and developers of  other network analysis systems that are similar
to Ondex in their intended usage.  From the application of this
methodology to Ondex, and learning by the experience of its usage in
the last 5 years, we have highlighted a set of issues that are likely
to be found in other network analysis systems when these are ported to
the Semantic Web.
To our knowledge this is the first time that a systematic assessment of the semantic mismatch between the data model and usage of pre-existing graph-based analysis tools and those of the Semantic Web is attempted. 

\subsection{The Ondex data structure, Life Sciences, and the Semantic Web}

Life Sciences data is characterized by its complexity, its high
interrelatedness, its heterogeneity, and by a multitude of naming and
identity issues \cite{nation,karp95, davidson95}.  Graph based models
are a natural fit, as they are in many disciplines, to deal with these
problems. From metabolic pathways to ecosystems to anatomies, graphs are
a convenient means to capture these relationships. Data in many forms can be
represented as a graph and the schema-less approach adopted in RDF and
other such representations affords a means of integrating data through
connections that can be as strong or weak as the applications require.

Ondex has taken this approach by providing a graph based model, a
collection of parsers that transform various bioinformatics resources
into its graph representation and plugin modules to perform further data
reduction and analysis.  The graph model of Ondex uses nodes to
represent concepts and edges to represent relationships between
them. Both concepts and entities can be characterized via a type (that
can be organized in an simple ontology), via an arbitrary set of
attributes, and via a set of predefined attributes which support the
representation of identifiers, information provenance, evidence and
context.

Thus Ondex's data model has an intuitive correspondence to that used
by RDF, and its parsers are \emph{de facto} equivalent to mappers
from the original resources to RDF.

There are, however, aspects of the Ondex representation, and its
usage, that make its use on the Semantic Web not as simple as it might
be. For instance its development has led to inconsistencies in the
interpretation of how the Ondex parsers transform the data into the
graph, one notable example being different interpretation of
\emph{provenance} information.

Our goals, therefore, are two-fold: First to develop a
normative Ondex data model and map its transformation to RDF; and second
to be able to describe a normative model against which the builders of
Ondex parser can model their transformations.  Both objectives are
relevant to adaptation of other network based analysis tools to the
Semantic Web.

\section{Semantic Analysis}

We have developed a simple methodology to examine the semantics of a graph based data integration and analysis tool, and to guide the transformation of its representation and usage as to be suitable for the Semantic Web.
In this methodology, we first list all elements that make up the data structure of the system, and for each of these elements, we execute a set of steps, which result in a document, for each element, which describes in natural language: the intended semantics of the element, its actual semantics in current practice, a definition, recommendations for best practices and recommendations for future developments. 
The documents are then circulated among stakeholders for comments and iterative refinements of the proposal.
We illustrate the steps that compose this methodology using its application on Ondex as an example, and in particular focusing on the \emph{CV} data element.
\subsection{Methodology}

\subsubsection{Listing data structure elements}
We first list the elements that make up the data structure of the system. In
the case of Ondex, these are \emph{Concept}, \emph{Relation},
\emph{ConceptClass}, \emph{RelationType}, \emph{Generalized Data Set},
\emph{CV} (Controlled Vocabulary), \emph{Accession} (identifier),
\emph{EvidenceType}, \emph{Context} (an extensive explanation of the
Ondex data types can be found in the detail of our analysis, available
at \cite{OndexSemDocs}).  
In the remainder of this example we will focus on \emph{CV}.

\subsubsection{Definition of the intended semantics}
We first elaborate a concise, informal definition of the intended
semantics of a data structure (e.g.: \emph{CV}). This definition is
based on the answers to two questions:``what is a \emph{CV} ?'' and
``when do you use a \emph{CV} ?''.  Answers to these question are derived
from the official documentation and from interviews with interested
parties (developers and core users).  We enrich this definition with a
few examples of values that are assigned to this data structure.  The
application of this step results, in our example, in :
\begin{itemize}
\item \emph{Definition}: Describes the bioinformatics origin of
Concept or accessions of Concepts (It is intended to represent
provenance information);
\item \emph{Examples}: ÒUNIGENEÓ, ÒGOÓ, ÒunknownÓ, ÒAFFYMETRIXÓ,
ÒBROADÓ, ÒNWBÓ (these are all examples of which are assigned to \emph{CV}).
\end{itemize}

\subsubsection{Observation of the actual usage}
We then analyze the actual usage of the data structure element, with
the help of a domain expert, to find patterns in the attribution of
values to the data structure, values that are inconsistent with its
intended usage, and degenerate usages. We compile a set of
observations for each of the patterns, inconsistencies or degenerate usages
that are found.

In the case of Ondex, and for other platform that present a plugin
architecture, the inspection of the code of plugins provides an easy
way to perform such analysis.  Some of the observations found for
\emph{CV} are:
 
\begin{itemize}
\item \emph{CV} is often in association with an
identifier (\emph{Accession} in the Ondex data structure) to
characterize its scope. This happens both in parsers that extract
information from ontologies such as the Gene Ontology (e.g.: \emph{CV}=GO) or some databases such as the the Unigene DNA sequence database
(e.g: \emph{CV}=Unigene) and plugins that perform mapping
operations. 

\item 
sometimes when \emph{CV} is associated to a \emph{Concept}, it
is assigned values that refer to the database from which the
information was extracted, rather than to the domain of identifiers for the concepts
in the database. This is for instance the case for the parser for the ATRegNet database of plant transcription factors.
(e.g.: \emph{CV}=``ATReg-Net'').

\item 
\emph{CV} is sometimes assigned values that indicate the format
from which some information was extracted, such as in Network Workbench (NWB) format
(e.g. \emph{CV}=``NWB'').

\item \emph{CV} is assigned an arbitrary identifier in plugins that
need to distinguish between different graphs.
\end{itemize}

\subsubsection{Analysis of the actual usage and normative definition}
Following the observations in the previous step, we elaborate a second
concise definition for the semantics of the data element, on which we
base the development of recommendations in the following steps. This
definition also traces the relations between the data structure
analyzed and elements of RDF that it more closely represents.  In the
case of \emph{CV}:

CV is currently used with several distinct meanings:
\begin{itemize}
\item when used in association with an identifier, \emph{CV} has the
meaning of a namespace, and characterizes the scope of the identifier.
\item when used with a \emph{Concept}, \emph{CV} has the meaning of provenance.
\end{itemize}

\subsubsection{Recommendation for best practice}
We elaborate a set of best practices, that are intended to restrict
the possible usage of the data structure to keep it coherent between
users and with a Semantic Web representation. Best practices are
designed to not require any change in the code base, and take into
account the observations previously derived. In our example:
\begin{itemize}

\item Usage of \emph{CV} in association to an \emph{Accession}:\\
When used as a namespace, \emph{CV} should be assigned values that
correspond unambiguously to the resources that provide a definition
for the identifier.  A pair (\emph{CV}, \emph{accession}) should be
semantically equivalent to a URI. In particular the following usage
should be avoided: \emph{CV} that are not specific enough, for
instance that correspond to a family of ontologies (e.g. OBO) rather
then a single ontology, to which identifiers are specific (e.g.; GO),
\emph{CV} that correspond to a technology used to generate data
(e.g.: Affymetrix) or to the institute providing the data (e.g.:
Broad).

\item Usage of \emph{CV} in association with a \emph{Concept}:\\ When
\emph{CV} is used to represent information about provenance, it is
intended to indicate the last source that asserted this information. In
the case of information originating from a database, \emph{CV} is intended as
the most specific authority that is responsible for the validity of
the data (this is often the last data source from which this concept
is derived).

\item Any other usage of CV is discouraged.

\end{itemize}

\subsubsection{Recommendation for future development}
We then present recommendations for future evolutions of the data
structure, that would help in enforcing the best practices and would
enable further integration with RDF:

\begin{itemize}
\item \emph{CV} should be split into two distinct elements,
corresponding to the meaning of ``Namespace'' and ``Provenance''.

\item Values for the ``Namespace'' element should be associated with one or more
effective namespaces that may be used to generate common URIs for the
concept.

\end{itemize}

\subsubsection{Request for comments}
Finally, all the specifications produced for the data structure element
are circulated to interested parties for feedback, which can lead to
new observations and further refinement.

\section{Results}

The analysis that we have conducted on the Ondex data structure
definition and usage highlights a series of issues that are not
limited to this platform, as they relate to typical assumptions behind
the usage of simple network based analysis platforms, and their
incongruence with a Semantic Web based representation of information.
We list here the most relevant issues we have found, with a brief
discussion of the risk they pose to make a consistent usage of network
based analysis tools on Semantic Web knowledge bases.

\subsection{Scope of information}
In Ondex, a \emph{Concept} (the equivalent of a resource in RDF) has
an identifier that is an integer generated when a graph is imported
into the system. A similar behavior can be found in Cytoscape. Both
Ondex and Cytoscape support the annotation of a Concept (or
node in the Cytoscape terminology) with identifiers, that can then be
used to derive identities between Concepts in different graphs (or in
different versions of the same graph).  This is typical of a
procedural, document based, data integration strategy where the
`document' provides an implicit scope for the validity of the
information that it represents.

In a web based context, it is important to explicitly define the scope
of validity of identifiers of resources and of the information
relative to these resources.  This is because in a distributed web environment,
it is not possible to import all the information before being able to
`name' and `access' the entities included.

In the Semantic Web framework, URIs act as identifiers with a global
scope that allow direct access to the relative information.  It is
also important to explicitly define the context of validity of
information, as the implicit context provided by a document has a
limited validity in a web framework, where information can easily be
filtered and recombined.

\subsection{Information basis}
When using a graph based data integration and analysis platform, it is
a tempting practice to use the graph for all information, without
making a distinction between the different basis for particular types
of information.  For instance it is common practice in an Ondex plugin
to represent, in the same graph, information that is based on `knowledge external to the system', information that is based on the
results of an analysis of the graph (e.g.: measures of betweenness and
centrality of nodes) and sometimes information that is based on the
specific instance of the platform (e.g.: graph coordinates for a given
layout).  This happens despite Ondex providing support for typing
concepts and relations via a simple ontology definition. Other
platforms are, in general, even more vulnerable to this `congestion'
of the graph.
This usage of the graph
data structure is acceptable in a procedural framework, where there is
a starting point that holds only `knowledge external to the system',
that is replicated in the system and never altered in its original
representation, and where information later added to the graph have
the implicit scope of the task that is being carried out.  In a web
based framework, however, it is necessary to distinguish information
that persists beyond the specific task carried out, information that
is dependent on a specific subset of information (i.e.: it is
invalidated when this subset is altered) and information that is not
shared, but specific to a given instance of execution of a tool.

\subsection{Cardinalities of relations}
Most network-based analysis tools, including Ondex, apply to the
network a data modeling approach that is typical of object oriented
(or framework based) systems and that is not consistent with a web
based representation of information.
This is particularly evident in the
case of relations. A tool like Ondex (or Cytoscape) will expect that,
if for a given concept the same property is asserted twice, with two
different values, the second value for this property will override the
first.  This is in contrast with a web based representation of
information where there is no limit on the number of values that a
property can be assigned for a given resource.

\subsection{Objectification of entities}
Another inconsistency that arises when an `object oriented' paradigm
is applied to Semantic Web resources derive from the fact that, when
entities are represented via objects, there is an additional entity
(the object) that has its own identifiers (the pointer). This can have
subtle consequences, in particular for the implementation of graph manipulation plugins.  For instance, a plugin can refer to the `first' or
in general to the `n-th' property asserted on a concept, simply by
retaining its pointer, and it can base its computations on this
ordering. There is not an equivalent of the `first' or `n-th' property
asserted, in the Semantic Web framework.

\subsection{Datatypes}
While some platform such as Cytoscape limit datatypes to a limited set
of basic types (strings, integers, booleans), platforms like Ondex
allow datatypes of arbitrary complexity (in practice, they allow
serializations of Java objects). This can limit
interoperability of systems for two reasons. First, other systems may
not be able to reconstruct an arbitrary Java object. Second, and more
importantly, data types are semantically opaque: complex datatypes
provide information without an explicit characterization of its
meaning.

\subsection{Over-specification}
Finally, we have observed that much imprecision stems from an
over-specification of the data-structure. In order to cope with characteristics as `provenance' and `evidence' of information, often the data structure require information, which cannot be guaranteed to be meaningful for the heterogenous nature of data on the web.
For instance Ondex requires information on
provenance and evidence for all \emph{concepts} and \emph{relations},
where provenance is intended to characterize the source of data (see
discussion on \emph{CV} in the previous section) and evidence its
validity.  Clear definitions for provenance and evidence apply to only
a subset of the information that can be represented in Ondex. For
instance, what is the evidence of an ontology term ? Or what is the
provenance of a value that is the result of a numerical analysis ?
Furthermore, users may simply not know the original data sources in
the detail that is necessary to assign correct evidence and provenance
information.  The result is a set of uninformative entries, ranging
from the generic ``imported from microarray-database'' to the ambiguous
``unknown''.

\section{Conclusions}

We have developed a simple, practical methodology to assess how the
documented semantics of a data integration tool differs from its
actual usage and, more specifically, where the semantic definition of
the data structures of these tools is underspecified to cope with
distributed information on the web.  While this simple method has been
devised to support the integration of Semantic Web functionalities in
Ondex, it describes a general approach that can be of help to the
adaptation of a variety of similar network based analysis tools to
operate on the Semantic Web.

Ondex exhibits problems of systems that have grown in an \emph{ad hoc}
manner that have under-specified semantics and roles for their data
models. The result are graphs that themselves have barriers to
integration. Our simple, practical approach to normalising the
project's understanding of its own data-model will have obvious
benefits within Ondex. 
As a preliminary result, it has enabled us to define a mapping between a subset of the Ondex data-model and that of the Semantic Web. This has been the basis for the development of an Ondex prototype which can consume and produce information in RDF.
Within the Ondex experience is a simple message that just
creating a graph does not mean integration; a common integration
pattern must be used. Otherwise, we have integration of format that is
still difficult to use.

Learning from the Ondex experience, we have identified problems that
are common to similar tools. We hope that this experience will help to
improve the information design of the next generation of data
integration, analysis and visualization platforms that will help in
fulfilling the promises of the Semantic Web.

\section*{Acknowledgment}

The authors gratefully acknowledge the UK Biotechnology and Biological Sciences Research Council (BBSRC) for funding this work under the Systems Approach to Biological Research (SABR) initiative (Grants: AS and CJR, BB/F006039/1; RS, BB/F006012/1; PL, BB/F006063/1). SCK was funded by a SABR project studentship.  Rothamsted Research is in receipt of grant in aid from the BBSRC.



\bibliographystyle{IEEEtran}
%

\end{document}